# Leaving the Fullerene Road: Presence and Stability of sp Chains in sp$^2$ Carbon Clusters and Cluster-Assembled Solids.

M. Bogana[1], L. Ravagnan[2], C.S. Casari[1], A. Zivelonghi[1], A. Baserga[1], A. Li Bassi[1], C.E. Bottani[1], S. Vinati[2], E. Salis[1,2], P. Piseri[2], E. Barborini[2], L. Colombo[3], P. Milani[2, *]

[1]NEMAS (Center of Excellence for NanoEngineered Materials and Surfaces), INFM-Dipartimento Ingegneria Nucleare, Politecnico di Milano, via Ponzio 34/3, 20133 Milano, Italy.

[2]Centro Interdisciplinare Materiali e Interfacce Nanostrutturati, INFM-Dipartimento di Fisica, Universita' di Milano, Via Celoria 16, 20133 Milano, Italy

[3]INFM-DEMOCRITOS National Simulation Center and Department of Physics, University of Cagliari, Cittadella Universitaria, 09042 Monserrato (Italy)

Abstract

We report the experimental and theoretical investigation of the growth and of the structure of large carbon clusters produced in a supersonic expansion by a pulsed microplasma source. The absence of a significant thermal annealing during the cluster growth causes the formation of disordered structures where sp$^2$ and sp hybridizations coexist for particles larger than roughly 90 atoms. Among different structures we recognize sp$^2$ closed networks encaging sp chains. This "nutshell" configuration can prevent the fragmentation of sp species upon deposition of the clusters thus allowing the formation of nanostructured films containing carbynoid species, as shown by Raman spectroscopy. Atomistic simulations confirm that the observed Raman spectra are the signature of the sp/sp$^2$ hybridization characteristic of the isolated clusters and surviving in the film and provide information about the structure of the sp chains. Endohedral sp chains in sp$^2$ cages represent a novel way in which carbon nanostructures may be organized with potential interesting functional properties.

*corresponding author: pmilani@mi.infn.it

The organization of carbon atoms from a plasma or a vapour into ordered structures and different allotropes is a matter of intense investigation. A paradigm of the carbon ability to provide fascinating structures is the efficient formation of $C_{60}$ or nanotubes by laser vaporization or arc discharge from graphite [1-5]. This aspect has stimulated an enormous experimental and theoretical work towards the identification of building blocks of $C_{60}$ and other carbon nanostructures and their assembling mechanisms [2].

Although a general and comprehensive understanding of the assembling of carbon nanostructures is still lacking, it is generally accepted that linear carbon sp chains are essential ingredients for the formation of carbon fullerenes and nanotubes [6, 7]. In particular, while calculations show that carbon chains are stable in high temperature carbon plasma [8], experiments prove that small sp clusters are found in the mass spectra of laser vaporized graphite [9]. As the number of atoms per cluster increases, $sp^2$ hybridization becomes energetically favoured and sp chains rearrange to form graphitic networks [3, 7]. The critical mass for sp-$sp^2$ rehybridization seems to depend upon the thermodynamic conditions and, in any case, it should not exceed few tenth of atoms.

Thermal annealing has been identified as a fundamental ingredient to transform sp structures into fullerenes and nanotubes and to collapse disordered carbon clusters into ordered structures [5, 10, 11]. The yield of fullerenes produced by laser vaporization in a room temperature buffer gas is almost negligible, whereas the use of a buffer gas at 1200 °C produces a yield of 30% [3, 11].

Exploring the "fullerene road" [2] from small sp chains to fullerenes, one encounters a class of loosely-defined objects named "non-fullerene clusters" [3]. They are considered as highly-reactive, metastable structures, evolving to fullerene cages under annealing [2, 3, 6, 11]. Since they are considered as unstable, non-fullerene clusters have never been considered as building blocks for solid-state carbon structures integrally formed by (or containing) sp species, like, for example, carbyne or graphine [12-14].

Recently it has been proved that a pure carbon solid containing sp structures can be produced by carbon cluster deposition in ultra high vacuum and the sp component can be identified and characterized by a precise Raman signature [15-17]. The production of a pure sp/$sp^2$ carbon solid by assembling carbon nanoparticles raises several questions which define the motivation of the present work: which is the structure of carbon clusters assembled to form a pure carbon solid containing sp chains? What is the microscopic structure responsible for the observed sp Raman signature? How reactive objects containing sp chains survive landing and how are they stabilized against cross-linking reactions? Is this an indication that sp-bonded allotropic carbon forms can be assembled by using suitable precursors?

In order to address the above questions, we have investigated the formation process and the resulting structure of carbon clusters produced in a pulsed microplasma cluster source (PMCS) [18], where thermal annealing is very much reduced compared to other methods. The key feature of the present work is the combined use of experiments and computer simulations, specifically designed to mimic real laboratory conditions. In particular, the present simulations are addressed to investigate non-fullerene clusters, as observed when thermal annealing is negligible.

We will show by real and virtual experiments that in a PMCS the collapse of clusters towards a perfect fullerene structure is not favoured. Large carbon clusters show the tendency to assume a disordered cage-like structure characterized by the coexistence of sp and $sp^2$ hybridization. By investigating the Raman spectra of cluster-assembled films, we prove that these structures survive the cluster deposition and are eventually incorporated in the growing films.

Carbon clusters in a supersonic expansion are produced by a PMCS as described in detail in ref. [18]. Briefly a graphite target is sputtered by an aerodynamically confined helium plasma [19]. The sputtering process transfers a reduced amount of energy to the extracted atoms and the background gas compared to laser vaporization [20]. The condensation regime in a sputtering source is similar to a gas aggregation source without interaction with a hot plasma plume [21]. Moreover, in the PMCS, the temperature of the buffer gas during the cluster residence in the PMCS prior to extractionis in a range of 100 K to 300 K so that no annealing of the clusters is taking place even for residence times of milliseconds [19, 22]. Nanoparticles evolve under different regimes: after the nucleation in the high temperature and high pressure region near the target, they grow by coagulation, as in aerosols [19].

In Fig.1 we show a typical mass spectrum produced by the PMCS. Large particles are observed whereas no small clusters roughly below $C_{60}$ are detectable. No prominent masses or magic numbers are present. The inset shows that odd clusters are also produced with a non-negligible abundance compared to the even ones. Small odd clusters in a mass region roughly lower than 30 atoms per particle are reported by several authors and are considered to be sp chains or rings [2, 9]. For larger clusters a reorganization to $sp^2$ cage structures with an even number of atoms should take place upon annealing [2].

It is very difficult to identify the structure of free clusters especially without the help of magic numbers. Gas phase ion chromatography has shown that clusters up to hundreds atoms can have, under particular conditions, different isomeric forms consisting of fullerene cages with short chains attached in stick and handle configurations [6]. These are considered as metastable structures and it is generally assumed that large carbon clusters have a fullerene-like structure with $sp^2$ bonding [2, 6]. The presence of very large odd clusters has never been reported in literature and it suggests that the clusters produced in the PMCS are different from those produced under significant thermal annealing. In our case the aggregation of sp small blocks seems to follow a different pathway and to produce non-fullerene isomers that do not evolve into pure $sp^2$ structures.

In order to achieve a deeper physical insight on the relevant atomic-scale mechanisms underlying the formation of large carbon clusters in a PMCS, we have performed tight-binding molecular dynamics (TBMD) simulations in the temperature and cluster-size conditions typical of the PCMS protocol. The adopted force model is derived by the TB Hamiltonian proposed by Xu et al. [23] which has been extensively used to investigate reactions involving fullerene-like molecules [24]. We have simulated the process of cluster aggregation by annealing at moderately high temperature (2000-3500 K) several high-density (in between 1.8 and 2.5 g cm$^{-3}$) carbon droplets containing from 60 up to 240 atoms. These parameters have been selected in order to mimic the conditions in the PMCS during the ablation of the target. No educated guesses have been adopted to get the actual atomic structure of the clusters, which was in fact obtained by a very long (up to millions time-steps) thermal annealings [25]. Atomic trajectories have been generated by velocity-Verlet algorithm using a time-step as short as 0.5ps.

Following the procedure described above, we have observed the formation of a rich variety of different isomers for any given cluster size. Clusters are characterized by highly-defective cage-like structures with both $sp^2$ and sp hybridisations (Fig. 2). Clusters in the range of roughly 60-70 atoms evolve towards a "disordered" fullerene-like structure as observed in Fig 2a. Clusters up to roughly 90 atoms typically show closed shells with 5-, 6- and 7-atoms rings as reported in the case of $C_{92}$ in Fig. 2b. For larger systems we observed a very interesting feature, namely the formation of complex structures characterized by a disordered $sp^2$ cage, trapping inside a network of complex sp or $sp^2$ substructures (Fig. 2c-h). Sometimes sp chains are bending outwards of the cage. Such external chains, however, have the tendency to fragment into $C_2$ dimers or other small molecules.

The key point of our simulations is that both even and odd clusters can be formed and that sp chain-like structures are indeed found to be stable inside cluster cages, thus providing a mixed sp/sp2 hybridization condition. This is shown in Fig. 3 where different time snapshots of the evolution of $C_{226}$ cluster are shown.

The presence of sp defects on the cage of large clusters (Fig. 2e) and the embedding of long sp chains inside the cage (Fig. 2c-h) can hardly be confirmed by direct experiments performed on the gas phase. Nevertheless, we here propose that the above peculiar cluster structure is responsible for the presence of sp species in the cluster-assembled films as reported in [15, 16]. In order to validate this guess, we have performed a direct comparison between the Raman spectra measured for films produced by depositing clusters generated in the PMCS and the Raman spectra calculated on simulated structures as reported in Figs. 3 and in the inset of Fig. 4.

The theoretical Raman spectra have been obtained by hierarchically combining TBMD and *ab initio* tools. At first TBMD simulations are used as scouts to search for equilibrium configurations, from which interesting sp chain-like objects have been extracted. The geometry of such fragments (decorated by hydrogen atoms so to saturate dangling bonds) has been fully optimised [26] by using the Gaussian03 package where the hybrid B3LYP density functional has been used together with standard Pople's 6-31G* basis set [27]. The same package has been used to evaluate Raman spectra. The comparison between experimental spectra and simulation is reported in Fig. 4. The experimental peak at 2100 cm$^{-1}$ is the signature of the presence of sp bonds [12, 15, 16], however it does not provide direct information about the structure and dimensions of the chains [12].

Our simulation shows that complex sp/sp$^2$ topologies sitting inside or on the surface of large carbon clusters (see the inset of Fig. 4 where the $C_{162}$ and $C_{113}$ are reported in inset a) and c) whereas the endohedral structure of $C_{162}$ and the surface structure of $C_{113}$ are reported in inset b) and d)). To calculate the Raman response of these structures we have saturated their dangling bonds with hydrogen (in green in Fig. 4) and fully optimized their geometry. We obtain a high density of active Raman modes around 2100 cm$^{-1}$ (6 modes between 2000 cm$^{-1}$ and 2200 cm$^{-1}$), with a dispersion that may explain the large width (more than 350cm$^{-1}$) of the experimentally observed peak. (Fig. 4). The good agreement between experiment and simulations suggest that the structures with the structure described above can be responsible for the observed Raman spectra.

In summary we have shown that carbon cluster production pathways without thermal annealing produce even and odd large particles where sp chains can coexist with sp$^2$ networks. In particular the structure of those clusters where sp chains are inside graphitic cages can explain why fragile carbynoid species can survive upon landing with kinetic energies per atom well above the sp stability [15, 16]. The sp$^2$ cages can act as protective shells absorbing efficiently the energy released by the impact. Moreover the sp chains are isolated in the shell so that cross-linking reactions are less favoured. Sp chains without protection rapidly transform into sp$^2$ species [17].

Our results provide an insight of the mechanisms underlying the formation of carbon materials containing carbynoid species and also demonstrate that these species embedded in sp$^2$ clusters can survive in a wider range of conditions compared to naked sp chains. Comparison of Raman spectra with atomistic simulation provide an insight of the structures that sp species assume in the nanostructured films.

The structure of non-fullerene clusters where sp and $sp^2$ hybridization coexists is somehow unexpected and very interesting from the topological point of view. In particular $sp^2$ cages with endohedral sp chains may be considered as a novel form of carbon and one can speculate about the fascinating electronic and magnetic properties of graphitic "nutshells" containing sp chains.

We thank Prof. W. Kraetschmer for many insightful discussions and Prof. G. Pacchioni for his help in the use of the Gaussian code. This work has been supported by MIUR under FIRB project "Carbon micro- and nanostructures".

FIGURE CAPTIONS

Figure 1. Detail of a time-of-flight mass spectrum of carbon clusters in a supersonic beam produced by the PMCS. No magic numbers are observed and clusters below $C_{60}$ are almost absent. The mass distribution is characterized by the presence of large even and odd clusters (inset). Typically the center of mass of the distribution is peaked around 900 atoms/cluster with a log-normal shape and geometrical standard deviation $\sigma_m=1.9$.

Figure 2. (coloured on line)
Cluster structures after 100 ps annealing at 3500K of droplets formed by: (**a**) 60 atoms, (**b,c,d,e**) 120 atoms, (**f,g**) 180 atoms and (**h**) 240 atoms. Atomic chains and *internal-cluster* structures have been high lightened in red and their dimension has been reported below. No educated guess has been used during the annealing procedure. Atoms forming chains are represented in red. Note that the cluster with 60 atoms does not show a fullerene structure.

Figure 3 (coloured on line)
Snapshots of cluster $C_{226}$ during annealing at 3500K at different at different time steps. The atoms forming the front portion of the cage have been removed in order to show more clearly the endohedral atoms. The number of atoms forming the internal structure represented in red is reported on the bottom right corner of each snapshot. We observe that the endohedral clusters from an initial sp2 structure evolve towards a linear structure characterized by sp hybridization which is stable continuing the annealing.

Figure 4 (coloured on line).
Comparison between experimental in situ Raman spectrum (black dots) taken from a cluster-assembled film deposited on a substrate at 100 K and the theoretical spectrum obtained by means of *ab initio* calculations (see text). The structures used for the simulations are reported in the inset. They are an endohedral sp structure from $C_{162}$ and a structure of the cage of $C_{113}$ respectively. The contributions from the endohedral structure of $C_{162}$ are in blue and those from $C_{133}$ are in red. Calculated Raman-shifts have been plotted rescaled of 3% for direct confrontation with experiment.

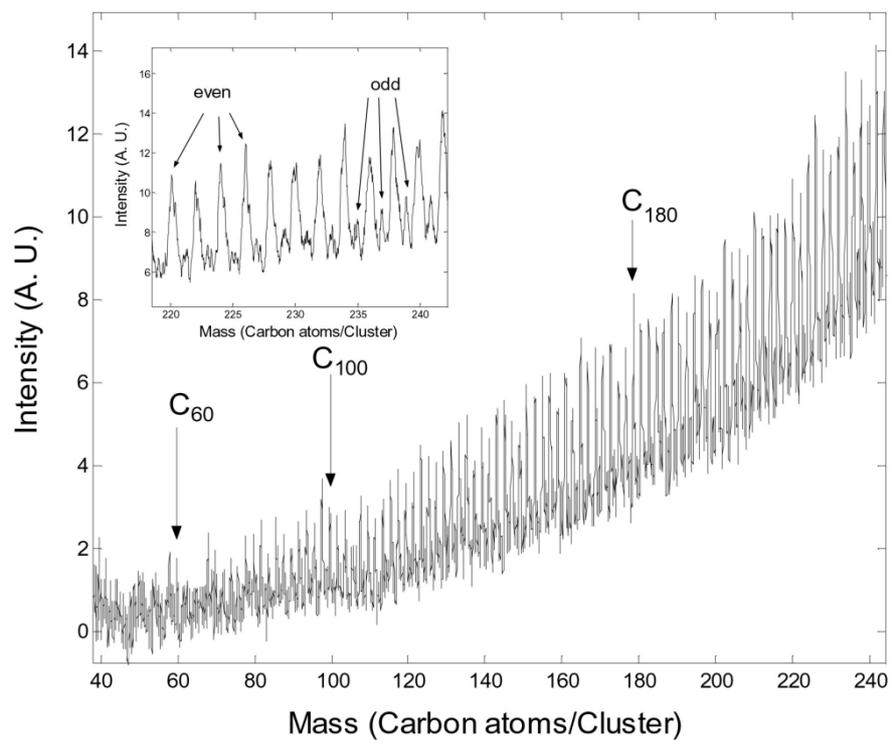

FIGURE 1

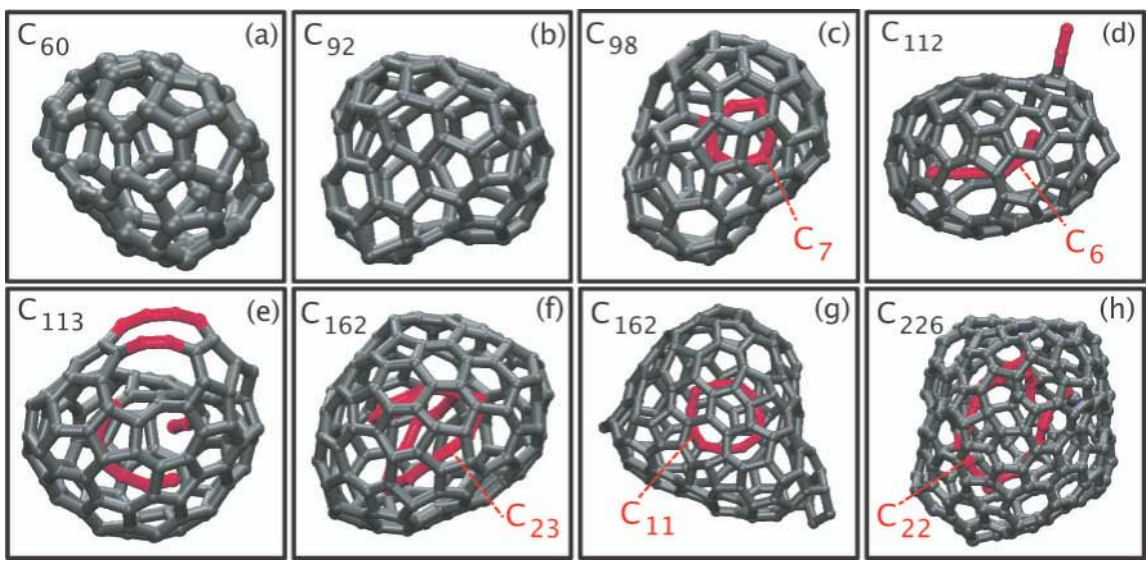

FIGURE 2

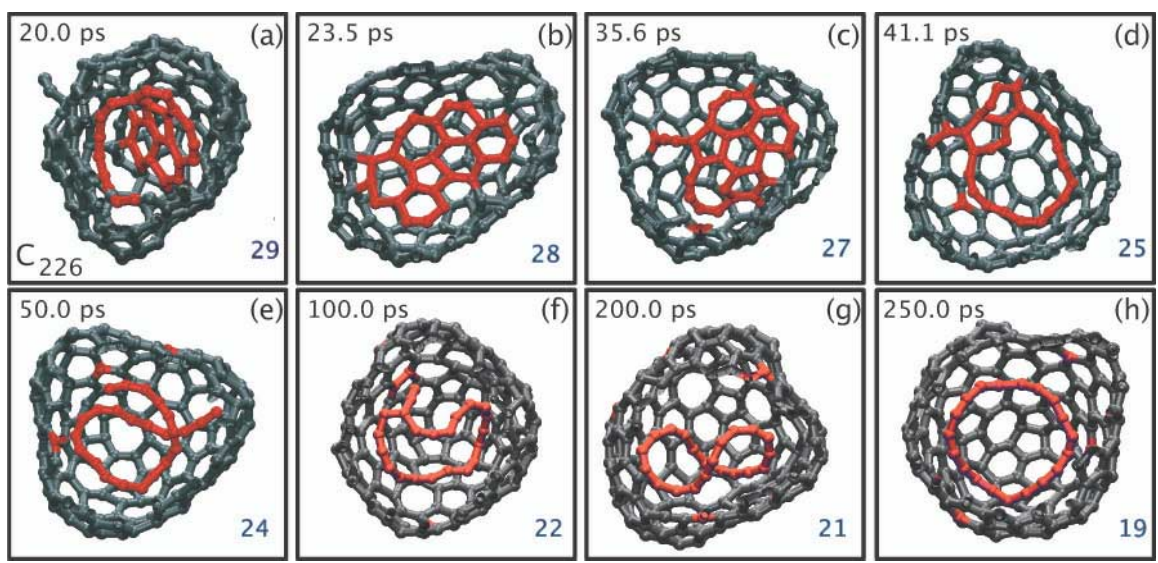

FIGURE 3

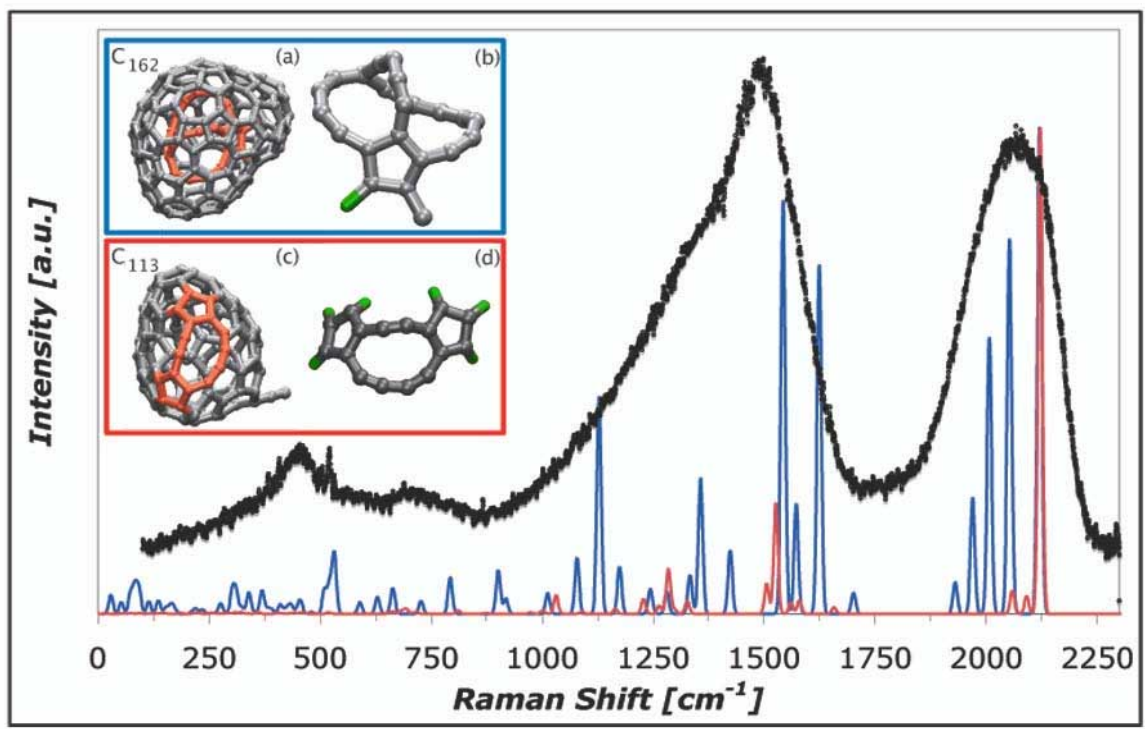

FIGURE 4